\documentclass[12pt]{article}
\usepackage{amssymb}
\usepackage{epsfig}
\begin{document}
%
\setlength{\baselineskip}{0.65cm}
\setlength{\parskip}{0.35cm}
%
\begin{titlepage}
%
\begin{flushright}
BNL-NT-03/4 \\
DO-TH 03/04 \\
RBRC-318 \\
March 2003
\end{flushright}

\vspace*{1.1cm}
\begin{center}
\LARGE

{\bf {Next-to-leading order QCD corrections}}\\

\medskip
{\bf {to ${\mathbf A}_{\mathrm{\mathbf{TT}}}$ for prompt photon production}}\\

\vspace*{1.5cm}
\large 
{A.\ Mukherjee$^{a}$, M.\ Stratmann$^{b}$, 
and W.\ Vogelsang$^{c,d}$}

\vspace*{1.0cm}
\normalsize
{\em $^a$Institut f{\"u}r Physik, Universit{\"a}t Dortmund,\\
D-44221 Dortmund, Germany}\\

\vspace*{0.5cm}
{\em $^b$Institut f{\"u}r Theoretische Physik, Universit{\"a}t Regensburg,\\
D-93040 Regensburg, Germany}\\

\vspace*{0.5cm}
{\em $^c$Physics Department, Brookhaven National Laboratory,\\
Upton, New York 11973, U.S.A.}\\

\vspace*{0.5cm}
{\em $^d$RIKEN-BNL Research Center, Bldg. 510a, Brookhaven 
National Laboratory, \\
Upton, New York 11973 -- 5000, U.S.A.}
\end{center}

\vspace*{1.5cm}
\begin{abstract}
      
We present a next-to-leading order QCD calculation of the cross section 
for isolated large-$p_T$ prompt photon production in collisions of 
transversely polarized protons. We devise a simple method of 
dealing with the phase space integrals in dimensional regularization
in the presence of the $\cos(2\Phi)$ azimuthal-angular dependence 
occurring for transverse polarization. Our results allow to calculate the 
double-spin asymmetry $A_{\mathrm{TT}}^{\gamma}$ for this process
at next-to-leading order accuracy, which may be used at
BNL-RHIC to measure the transversity parton distributions of 
the proton.
\end{abstract}
\end{titlepage}
\newpage

\section{Introduction}
%
\noindent
The partonic structure of spin-1/2 targets at the leading-twist level
is characterized entirely by the unpolarized, longitudinally polarized, 
and transversely polarized distribution functions $f$, $\Delta f$, 
and $\delta f$, respectively \cite{ref:jaffeji}. 
By virtue of the factorization theorem \cite{ref:fact}, these non-perturbative
parton densities can be probed universally in a multitude of 
inelastic scattering processes, for which it is possible
to separate (``factorize'') the long-distance physics relating
to nucleon structure from a partonic short-distance scattering
that is amenable to QCD perturbation theory.
Combined experimental and theoretical efforts have led to an improved
understanding of the spin structure of longitudinally polarized nucleons, 
$\Delta f$, in the past years. In contrast, the ``transversity'' 
distributions $\delta f$, first introduced in \cite{ref:ralston}, 
remain the quantities about which we have the least knowledge. 

Current and future experiments are designed to further unravel the
spin structure of both longitudinally and transversely polarized nucleons.
Information will soon be gathered for the first time from 
polarized proton-proton collisions at the BNL Relativistic Heavy
Ion Collider (RHIC) \cite{ref:rhic}. Collisions of transversely polarized 
protons will be studied, and the potential of RHIC in accessing transversity 
$\delta f$ in transverse double-spin asymmetries $A_{\mathrm{TT}}$ 
was recently examined in \cite{ref:attlo} for high transverse momentum 
$p_T$ prompt photon and jet production. Several other studies of 
$A_{\mathrm{TT}}$ for these reactions have been presented in the past
\cite{ref:attold,ref:artru,ref:jaffesaito}, as well as 
for the Drell-Yan process \cite{ref:ralston,ref:dylo,ref:dynlo,ref:wvkern}. 
With the exception of the latter reaction \cite{ref:dynlo,ref:wvkern},
all of these calculations were performed 
at the lowest order (LO) approximation only.
As is well known, next-to-leading order (NLO) QCD 
corrections are generally indispensable in order to arrive at a firmer 
theoretical prediction for hadronic cross sections and spin asymmetries.
Only with their knowledge can one reliably confront theory with experimental
data and achieve the goal of extracting information on the partonic spin 
structure of nucleons.

In this paper we extend the results of \cite{ref:attlo} for isolated
high-$p_T$ prompt photon production, $pp\to\gamma X$, to the NLO of QCD. 
Apart from the motivation given above, also interesting
new technical questions arise beyond the NLO in case of transverse
polarization. Unlike for longitudinally polarized cross sections where
the spin vectors are aligned with momentum, transverse spin vectors
specify extra spacial directions, giving rise to non-trivial
dependence of the cross section on the azimuthal angle of
the observed photon. As is well-known \cite{ref:ralston}, 
for $A_{\mathrm{TT}}$ this 
dependence is always of the form $\cos(2\Phi)$, if the $z$ axis is defined
by the direction of the initial protons in their center-of-mass
system (c.m.s.), and the spin vectors are 
taken to point in the $\pm x$ direction. Integration over 
the photon's azimuthal angle is therefore not appropriate. 
On the other hand, standard techniques developed in the literature 
for performing NLO phase-space integrations usually rely on the 
choice of particular reference frames that are related in complicated 
ways to the one just specified. This makes it difficult to 
fix $\Phi$ in the higher order phase space integration. The
problem actually becomes more severe if dimensional regularization 
techniques are used for dealing with the collinear and infrared
singularities, as is customary. Even for the kinematically 
rather simple Drell-Yan process the NLO calculation for the
cross section with transverse polarization is quite more 
complicated as for the unpolarized or longitudinally polarized
cases \cite{ref:dynlo}. In this paper, we will present 
a new general technique which facilitates NLO calculations with transverse
polarization by conveniently projecting on the azimuthal dependence 
of the matrix elements in a covariant way. This method then allows 
us to carry out phase space integrals with standard tools known from 
unpolarized calculations.

After presenting our technique and verifying that it recovers the known 
result for the transversely polarized NLO Drell-Yan
cross section, we apply it to high-$p_T$ prompt photon production.
We also present some first numerical calculations of the
cross sections and the transverse spin asymmetry for this process
at NLO. Here we of course have to rely on some model 
for the transversity densities, for which we make use of the 
Soffer inequality \cite{ref:soffer}. As in experiment, we 
impose an isolation cut on the photon. We find a moderate size
of the NLO corrections and the expected reduced scale dependence of the cross 
section at NLO.

\section{Calculation of the NLO Corrections}
%
\subsection{Preliminaries}
%
The transversity density $\delta f(x,\mu)$ is defined
\cite{ref:jaffeji,ref:ralston,ref:artru,ref:ratcliffe} as the difference of 
probabilities for finding a parton of flavor $f$ at scale $\mu$ and
light-cone momentum fraction $x$ with its spin aligned 
($\uparrow\uparrow$) or anti-aligned ($\downarrow\uparrow$)
to that of the transversely polarized nucleon:
\begin{equation}
\label{eq:pdf}
\delta f(x,\mu) \equiv f_{\uparrow\uparrow}(x,\mu) -
                       f_{\downarrow\uparrow}(x,\mu)
\end{equation}
(an arrow always denotes transverse polarization in the following).
The unpolarized densities are recovered by taking the sum in 
Eq.~(\ref{eq:pdf}). When the transverse polarization is
described as a superposition of helicity eigenstates, $\delta f$ 
reveals its helicity-flip, chirally odd,
nature \cite{ref:jaffeji,ref:artru}. 
As a result, there is no leading-twist transversity 
gluon density, since helicity changes by two units cannot 
be absorbed by a spin-1/2 target 
\cite{ref:jaffeji,ref:artru,ref:ji2}\footnote{
We note that a gluon density does contribute beyond leading 
twist \cite{ref:ji2,ref:st}, where it will lead
to terms in $A_{\mathrm{TT}}$ strongly suppressed by inverse powers of the 
photon $p_T$. An estimate of such effects could follow the lines
in \cite{ref:st}.}.
The property of helicity conservation in QCD hard scattering processes
implies that there have to be two soft hadronic pieces in the process 
that each flip chirality,
in order to give sensitivity to transversity. One possibility, 
which we are going to consider in the following, is to have two transversely 
polarized hadrons in the initial-state and to measure double-spin asymmetries 
\begin{equation}
\label{eq:att}
A_{\mathrm{TT}} = \frac{\frac{1}{2}
              \left[d\sigma(\uparrow\uparrow) - 
              d\sigma(\uparrow\downarrow)\right]}
                                  {\frac{1}{2}
                \left[d\sigma(\uparrow\uparrow) + 
               d\sigma(\uparrow\downarrow)\right]}
\equiv \frac{d\delta\sigma}{d\sigma}\;\;.
\end{equation}
Here $d\delta\sigma$ denotes the transversely polarized cross section.
$A_{\mathrm{TT}}$ is expected to be rather small for most processes 
\cite{ref:attold,ref:jaffesaito,ref:attlo}, since gluonic contributions
are absent in the numerator while in the denominator they
often play a dominant role\footnote{The only exception is the Drell-Yan 
process, which however suffers from rather low rates.}.   
Nevertheless, the LO study \cite{ref:attlo} suggests that the asymmetry for
prompt photon production should be measurable at RHIC, provided
the transversity densities are not too small.

According to the factorization theorem \cite{ref:fact} the fully
differential transversely polarized single-inclusive 
cross section $A+B\rightarrow \gamma+X$ for the production of a prompt photon 
with with transverse momentum $p_T$,
azimuthal angle $\Phi$ with respect to the initial spin axis, 
and pseudorapidity $\eta$ reads 
\begin{eqnarray} 
\label{eq:xsec}
\frac{d^3\delta \sigma}{dp_T d\eta d\Phi}
&=& \frac{p_T}{\pi S} \sum_{a,b}  
\int_{VW}^{V}\frac{dv}{v(1-v)} 
\int^1_{VW/v}\frac{dw}{w}\delta f_a(x_a,\mu_F) \delta f_b(x_b,\mu_F) 
\nonumber \\[3mm]
&\times&  \,\left[ 
\frac{d\delta \hat{\sigma}^{(0)}_{ab\to \gamma}(v)}{dvd\Phi} 
\delta (1-w) + \frac{\alpha_s(\mu_R)}{\pi} \, \frac{d\delta 
\hat{\sigma}^{(1)}_{ab\to\gamma}(s,v,w,\mu_R,\mu_F)}{dvdwd\Phi}
\right] \;\; , 
\end{eqnarray}
with hadron-level variables 
\begin{equation}
V\equiv 1+\frac{T}{S} \; \; , \;\;\;\; 
W\equiv \frac{-U}{S+T} \; \; , \;\;\;\; 
S\equiv (P_A+P_B)^2 \; \; , \;\;\;\; T\equiv (P_A-P_{\gamma})^2 
\; \; , \;\;\;\; 
U\equiv (P_B-P_{\gamma})^2 \;\; ,
\end{equation}
in obvious notation of the momenta, and corresponding partonic ones 
\begin{equation} \label{partvar}
v\equiv 1+\frac{t}{s} \; \; , \;\;\;\; 
w\equiv \frac{-u}{s+t} \; \; , \;\;\;\; 
s\equiv (p_a+p_b)^2 \; \; , \;\;\;\; t\equiv 
(p_a-p_{\gamma})^2 \; \; , \;\;\;\; u\equiv (p_b-p_{\gamma})^2 \;\; .
\end{equation}
Neglecting all masses, one has the relations
\begin{equation} \label{further}
s=x_a x_b S \;\; , \;\;\;\; t=x_a T \;\; , \;\;\;\; 
u=x_b U \;\; , \;\;\;
x_a = \frac{VW}{vw} \;\; , \;\;\;\; x_b = \frac{1-V}{1-v} \;\;.
\end{equation}
The $d\delta\hat{\sigma}^{(i)}_{ab\to\gamma}$ are the LO ($i=0$) and NLO 
($i=1$) contributions in the partonic cross sections for the reactions
$a b\rightarrow \gamma X$. 
$\mu_R$ and $\mu_F$ are the renormalization and factorization scales.

\subsection{Projection Technique for Azimuthal Dependence}

Let us consider the scattering in the hadronic c.m.s.\ frame, 
assuming both initial spin vectors to be in $\pm x$ direction.
Then, on general grounds, for a parity-conserving theory with 
vector couplings, the $\Phi$-dependence of the cross section
is constrained to be of the form $\cos (2\Phi)$:
\begin{equation} 
\label{eq2}
\frac{d^3\delta \sigma}{dp_T d\eta d\Phi}\;\equiv\;
\cos (2\Phi)\,\left\langle \frac{d^2\delta\sigma}{dp_T d\eta}\right\rangle
\; .
\end{equation} 
We may obtain $\left\langle d^2\delta\sigma/dp_T d\eta\right\rangle$
by integrating the cross section over $\Phi$ with a $\cos (2\Phi)$ weight:
\begin{equation} 
\label{eq3}
\left\langle \frac{d^2\delta\sigma}{dp_T d\eta}\right\rangle\;=\;
\frac{1}{\pi}
\int_0^{2\pi}d\Phi\cos (2\Phi)\;\frac{d^3\delta \sigma}{dp_T d\eta d\Phi}\; .
\end{equation} 
For the lowest order contribution to prompt-photon production
in Eq.~(\ref{eq:xsec}) one has only the channel $q\bar{q}\to\gamma g$.
Polarization for, say, the initial quark may be projected out by 
\begin{equation}
\label{eq:trspin}
u(p_a,s_a)\,\bar{u}(p_a,s_a)=\frac{1}{2}\slash{\!\!\!p}_a\left[
1+\gamma_5\slash{\!\!\!s}_a\right]\; ,
\end{equation}
where $p_a$ and $s_a$ are the quark's momentum and transverse spin vector,
and $u(p_a,s_a)$ its Dirac spinor. One readily finds for the LO process
\begin{equation} 
\label{eq4}
\left\langle \frac{d\delta \hat{\sigma}^{(0)}_{q\bar{q}\to \gamma g}(v)}{dv} 
\right\rangle\;=\;
\frac{2C_F}{N_C}\,\frac{\alpha\alpha_s}{s} e_q^2 \; ,
\end{equation} 
where $C_F=4/3$, $N_C=3$ and $e_q$ is the fractional quark charge.

As discussed in the Introduction, in the NLO calculation one
wants to make as much use as possible of calculational 
techniques established for the unpolarized case. For a single inclusive 
cross section such as prompt photon production, the appropriate methods 
were developed in \cite{ref:nlos}. They involve integration 
over azimuthal angles. We therefore would like to follow a projection 
analogous to Eq.~(\ref{eq3}); however, we should formulate it
in a covariant way. To this effect, we first note that 
the factor $\cos (2\Phi)/\pi$ in the cross section actually results
from the covariant expression
\begin{equation} 
\label{eq5}
{\cal F}(p_{\gamma},s_a,s_b)\;=\;
\frac{s}{\pi t u} 
\,\left[ 2 \,(p_{\gamma}\cdot s_a)\, (p_{\gamma}\cdot s_b)\; +\; 
\frac{t u}{s} \,(s_a \cdot s_b) \right] \;,
\end{equation} 
which reduces to $\cos (2\Phi)/\pi$ in the hadronic c.m.s.\ frame. We may,
therefore, use ${\cal F}(p_{\gamma},s_a,s_b)$ instead of the explicit 
$\cos (2\Phi)/\pi$. 

Even though employing ${\cal F}(p_{\gamma},s_a,s_b)$ becomes a real 
advantage only at NLO, let us illustrate its use in case of the LO cross 
section for the partonic reaction $q\bar{q}\to \gamma g$. We there have
\begin{equation} 
\label{eq6}
\frac{d\delta^2 \hat{\sigma}^{(0)}_{q\bar{q}\to \gamma g}}{dtd\Phi} \;=\; 
\frac{1}{32\pi^2 s^2}\;\delta |M(q\bar{q}\to \gamma g)|^2\; ,
\end{equation} 
where $\delta |M|^2$ is the squared invariant matrix element 
for the reaction with transverse polarization and reads:
\begin{equation} 
\label{eq7}
\delta |M(q\bar{q}\to \gamma g)|^2= (e e_q g)^2\,\frac{4C_F}{N_C}\, 
\frac{s}{tu}\,\left[ 2 \,(p_{\gamma}\cdot s_a)\, (p_{\gamma}\cdot s_b)\; +\; 
\frac{t u}{s} \,(s_a \cdot s_b) \right] \; .
\end{equation} 
One recognizes the factor ${\cal F}(p_{\gamma},s_a,s_b)$ emerging in 
$\delta |M|^2$. We now multiply $\delta |M|^2$ by 
${\cal F} (p_{\gamma},s_a,s_b)$, equivalent to the multiplication by
$\cos (2\Phi)/\pi$ in Eq.~(\ref{eq3}).  The resulting expression may then 
be integrated over the full azimuthal phase space without producing a 
vanishing result, unlike the case of $\delta |M|^2$ itself. 
This integration may again be performed in a covariant way by noting 
first that the dependence of ${\cal F}(p_{\gamma},s_a,s_b)\, \delta |M|^2$ on 
the spin vectors comes as $(p_{\gamma}\cdot s_a)^2(p_{\gamma}\cdot s_b)^2$,
$(p_{\gamma}\cdot s_a)(p_{\gamma}\cdot s_b)(s_a\cdot s_b)$, and
$(s_a\cdot s_b)^2$. The first two of these terms correspond 
to contractions with the tensors $p_{\gamma}^{\mu}p_{\gamma}^{\nu}
p_{\gamma}^{\rho}p_{\gamma}^{\sigma}$ and $p_{\gamma}^{\mu}p_{\gamma}^{\nu}$,
respectively. Expanding these tensors into all possible tensors made up of the
metric tensor and the incoming partonic momenta, one finds straightforwardly
\begin{eqnarray} 
\label{eq8}
&&\int d\Omega_{\gamma}\,(p_{\gamma}\cdot s_a)^2(p_{\gamma}\cdot s_b)^2 =
\int d\Omega_{\gamma}\, \frac{t^2u^2}{8s^2}
\left(2 (s_a\cdot s_b)^2+s_a^2 s_b^2\right)\;=\;
\int d\Omega_{\gamma}\, \frac{3t^2u^2}{8s^2} \; ,\nonumber \\
&&\int d\Omega_{\gamma}\, (p_{\gamma}\cdot s_a) 
(p_{\gamma}\cdot s_b)(s_a\cdot s_b) =
-\int d\Omega_{\gamma}\, \frac{tu}{2s}
(s_a\cdot s_b)^2\;=\;-\int d\Omega_{\gamma}\, \frac{tu}{2s} \; ,
\end{eqnarray} 
where $\int d\Omega_{\gamma}$ denotes integration over the
photon phase space, and where we have chosen both spin vectors to 
point in the same direction. We also recall that $s_i\cdot p_a=
s_i\cdot p_b=0$ ($i=a,b$) and $s_a^2=s_b^2=-1$. 
We emphasize that after the replacements (\ref{eq8}) the whole invariant 
phase space over $p_{\gamma}$ remains to be integrated, including
the (now trivial) azimuthal part, as indicated by the $\int\,d\Omega_{\gamma}$ 
on the right hand side. This is the virtue of our method that
becomes particularly convenient at NLO. It is crucial here that the other
observed (``fixed'') quantities, transverse momentum $p_T$ and rapidity 
$\eta$, are determined entirely by scalar products $(p_a\cdot
p_{\gamma})$ and $(p_b\cdot p_{\gamma})$. This 
allows the above tensor decomposition with tensors
only made up of $p_a$ and $p_b$ and of course the metric tensor.

Inserting all results, and including the azimuthal part of the
$d\Omega_{\gamma}$ integration, we find 
\begin{equation} 
\label{eq9}
\langle \delta |M(q\bar{q}\to \gamma g)|^2 
\rangle= (e e_q g)^2\,\frac{4C_F}{N_C}\; , 
\end{equation} 
and hence, using Eq.~(\ref{eq6}), we recover Eq.~(\ref{eq4}).

In the NLO calculation, one has $2\to 3$ reactions $
ab\to \gamma c d$. For an inclusive photon spectrum, one
integrates over the full phase spaces $d\Omega_c$ and
$d\Omega_d$ of particles $c$ and $d$, respectively. 
The momentum of particle $d$ may be fixed by momentum conservation,
and the integration is trivial. One then ends up with 
\begin{equation}
\label{eq10}
\int d\Omega_{\gamma}\int d\Omega_c\;{\cal F}(p_{\gamma},s_a,s_b)\,
\delta |M(ab\to \gamma c d)|^2\; .
\end{equation} 
Besides scalar products of the $s_i$ $(i=a,b)$ with $p_{\gamma}$, the
integrand may contain terms $\propto(s_a\cdot p_c)(s_b\cdot p_c)$ and 
$\propto(s_i\cdot p_c)$. As before, we may expand the
ensuing tensor and vector integrals in terms of the available
tensors. As far as the integration over $d\Omega_c$ is concerned, 
such tensors may be made up of the metric tensor, $p_a$, $p_b$, 
and $p_{\gamma}$. It is also important to keep in mind that in the NLO 
calculation we will need to use dimensional regularization
due to the presence of singularities in the phase space integrations.
We find in $d=4-2\varepsilon$ dimensions:
\begin{eqnarray} 
\label{eq11}
\int d\Omega_c\, (p_c\cdot s_a) (p_c\cdot s_b) &=& \int d\Omega_c\left\{
\frac{t u}{s} \left[ \frac{1}{2}{\cal A}-{\cal B}\right] (s_a\cdot s_b)+ 
\left[ (1-\varepsilon){\cal A}-{\cal B}
\right](p_{\gamma}\cdot s_a)(p_{\gamma}\cdot s_b) \right\}\; ,
\nonumber \\
\int d\Omega_c\, (p_c\cdot s_i)  &=& \int d\Omega_c\;{\cal C} 
\cdot (p_{\gamma}\cdot s_i)\; ,
\end{eqnarray} 
where
\begin{eqnarray} 
\label{eq12}
{\cal A} &=& \frac{2}{(1-2 \varepsilon)} \;{\cal C}^2 \; , \nonumber \\
{\cal B} &=& \frac{1}{(1-2 \varepsilon)}\;\frac{t_c u_c}{t u}\; , \nonumber \\
{\cal C} &=& -\frac{s s_{\gamma c} -t u_c-t_c u}{2 t u} \; ,
\end{eqnarray} 
with
\begin{equation}
\label{eq13}
t_c\equiv(p_a-p_c)^2\,\, ,\,\, u_c\equiv(p_b-p_c)^2\,\, ,\,\,
s_{\gamma c}\equiv(p_{\gamma}+p_c)^2 \; .
\end{equation}
After scalar products involving $p_c$ with the $s_i$ have
been eliminated in this way, only those with $(p_{\gamma}\cdot
s_i)$ remain. As in our LO example, when we apply the factor
${\cal F}(p_{\gamma},s_a,s_b)$, these terms enter as
$(p_{\gamma}\cdot s_a)^2(p_{\gamma}\cdot s_b)^2$ and 
$(p_{\gamma}\cdot s_a)(p_{\gamma}\cdot s_b)$. We then may
use Eq.~(\ref{eq8}) after appropriate modification to $d=4-2
\varepsilon$ dimensions:
\begin{eqnarray} 
\label{eq14}
\int d\Omega_{\gamma}\,(p_{\gamma}\cdot s_a)^2(p_{\gamma}\cdot s_b)^2 &=&
\int d\Omega_{\gamma}\,\frac{t^2u^2}{4(1-\varepsilon)(2-\varepsilon)s^2}
\left[2 (s_a\cdot s_b)^2+s_a^2 s_b^2
\right] \; , \nonumber \\
\int d\Omega_{\gamma}\,(p_{\gamma}\cdot s_a) (p_{\gamma}\cdot s_b)
(s_a\cdot s_b) &=&
-\int d\Omega_{\gamma}\,\frac{tu}{2(1-\varepsilon)s}(s_a\cdot s_b)^2 \; .
\end{eqnarray} 
After this step, there are no scalar products involving the $s_i$ left
in the squared matrix element (except the trivial $s_a\cdot s_b=-1$).
We may now integrate over all phase space, employing techniques
familiar from the corresponding calculations in the unpolarized
and longitudinally polarized cases. As a check, we have applied our 
method to the Drell-Yan transversity cross section and recovered the known 
NLO result \cite{ref:wvkern} in a straightforward manner. For the 
interested reader, we list some details of this calculation in 
the Appendix. 

\subsection{Details of the NLO Calculation for Prompt Photon Production}
From here on, all steps in the calculation are fairly standard,
albeit still involved and lengthy. Since many of them have
been documented in previous papers 
\cite{ref:nlos,ref:nloaur,ref:nlogv,ref:nlopion}, we only give a brief
summary here. We emphasize that the general method we have employed
is to perform the integrations over the phase space of the 
unobserved particles in the $2\to 3$ contributions 
analytically. We have also simultaneously calculated the unpolarized 
cross section and found agreement with the expressions available
in the literature \cite{ref:nloaur,ref:nlogv}.

At NLO, there are two subprocesses that contribute for transverse 
polarization:
\begin{eqnarray} 
\label{loproc}
q\bar{q}&\to& \gamma X \nonumber \; ,\\
qq&\to& \gamma X  \; .
\end{eqnarray}
The first one of course was already present at LO, where 
$X=g$. At NLO, one has virtual corrections to the Born
cross section ($X=g$), but also $2\to 3$ real emission
diagrams, with $X=gg+q\bar{q}+q'\bar{q}'$. For the second 
subprocess, $X=qq$.  All contributions are treated as discussed
in the previous subsection, i.e., we project on their 
$\cos(2\Phi)$ dependence by multiplying with the 
function ${\cal F}(p_{\gamma},s_a,s_b)$ in Eq.~(\ref{eq5})
and integrating over the azimuthal phase space using
Eqs.~(\ref{eq11}) and (\ref{eq14}).

Owing to the presence of ultraviolet, infrared, and
collinear singularities at intermediate stages of the
calculation, it is necessary to introduce a regularization.
Our choice is dimensional regularization, that is, the 
calculation is performed in $d=4-2\varepsilon$ space-time 
dimensions. Subtractions of singularities are made 
in the $\overline{\rm{MS}}$ scheme throughout. 

Projection on a definite polarization state for the initial
partons involves the Dirac matrix $\gamma_5$, as is
evident from Eq.~(\ref{eq:trspin}). It is well known that 
dimensional regularization becomes a somewhat subtle issue 
if $\gamma_5$ enters the calculation, the reason being that 
$\gamma_5$ is a genuinely four-dimensional object with 
no natural extension to $d\neq 4$ dimensions. 
Extending the relation $\{\gamma_5,\gamma^{\mu}\}=0$
to $d$ dimensions leads to algebraic inconsistencies 
in Dirac traces with an odd number of $\gamma_5$ \cite{ref:cfh}.
Owing to the chirally odd nature of transversity, 
in our calculation all Dirac traces contain two
$\gamma_5$ matrices, and there should be no
problem using a naive, totally anticommuting $\gamma_5$ in
$d$ dimensions. Nevertheless, we also did the 
calculation using the widely-used ``HVBM scheme''~\cite{ref:hvbm}
for $\gamma_5$, which is known to be fully consistent.
It is mainly characterized by splitting the 
$d$-dimensional metric tensor into a four-dimensional and a 
$(d-4)$-dimensional one. In the four-dimensional subspace, 
$\gamma_5$ continues to anti-commute with the other Dirac matrices; 
however, it commutes with them in the $(d-4)$-dimensional one. 
The HVBM scheme thus leads to a higher complexity of the 
algebra\footnote{We use the program {\sc Tracer}~\cite{ref:tracer} 
to perform Dirac traces in $d$ dimensions.}
and of phase space integrals. We found the same final
answers for both $\gamma_5$ prescriptions in all our calculations.

Ultraviolet poles in the virtual diagrams are removed by the 
renormalization of the strong coupling constant at a scale $\mu_R$.
Infrared singularities cancel in the sum between virtual
and real-emission diagrams. After this cancellation,
only collinear poles are left. These result for example from a 
parton in the initial state splitting collinearly into 
a pair of partons, corresponding to a long-distance contribution in the 
partonic cross section. From the factorization theorem it follows that 
such contributions need to be factored, at a factorization
scale $\mu_F$,  into the parton distribution 
functions. A similar situation occurs in the final-state. The
high-$p_T$ photon may result from collinear radiation
off a quark, which again is singular. This singularity is 
absorbed into a ``quark-to-photon'' fragmentation function 
\cite{ref:nloaur,ref:nlogv} that
describes photon production in jet fragmentation and hence
by itself contains long-distance information. The fragmentation 
contribution has not been written down in Eq.~(\ref{eq:xsec}). 
It has a structure similar to Eq.~(\ref{eq:xsec}), but with an 
extra integration over the fragmentation function. Its size 
also depends on the experimental selection of prompt photon events, 
as we will discuss below. 

The subtraction of initial-state collinear singularities is
particularly simple in case of transversity since there is
no gluon transversity and only $q\to qg$ collinear splittings
can occur. Only the process $q\bar{q}\to\gamma g g$ has
such poles. Their cancellation is effected by adding a 
``counterterm'' that has the structure (for radiation off
the initial quark)
\begin{equation}\label{subfac1}
-\frac{\alpha_s}{\pi} \, \int_0^1 dx\;\delta H_{qq}(x,\mu_F)\,
\frac{d\delta \hat{\sigma}^{(0)}_{q\bar{q}\to \gamma g}(x s,x t,u,
\varepsilon)}{dv}\,\delta \!\left( x\, (s+t)+u \right)  \;\; ,
\end{equation}
where in the $\overline{\rm{MS}}$ scheme
\begin{equation} \label{subfac}
\delta H_{qq} (z,\mu_F) \equiv \left(-\frac{1}{\varepsilon}
+\gamma_E-\ln 4\pi \right) \delta
P_{qq} (z) \left( \frac{s}{\mu_F^2} \right)^{\varepsilon} \; ,
\end{equation}
with the LO transversity splitting function \cite{ref:lokernel}
\begin{equation} \label{plo}
\delta P_{qq} (z) = C_F \left[ \frac{2z}{(1-z)_+}+\frac{3}{2}\delta
(1-z)\right] \; .
\end{equation}
Here the ``plus''-distribution is defined in the usual way.
As indicated in Eq.~(\ref{subfac1}), the $2\to 2$ cross section
in the integrand needs to be evaluated in $d$ dimensions. The
result, which turns out to be the same in the anticommuting $\gamma_5$
and the HVBM schemes, is given by 
\begin{equation}
\left\langle \frac{d\delta \hat{\sigma}^{(0)}_{q\bar{q}\to \gamma g}(s,t,u,
\varepsilon)}{dt}\right\rangle =
\frac{2 C_F}{N_C}\frac{\alpha\alpha_s}{s^2}e_q^2\;
\frac{\mu^{2\varepsilon}}{\Gamma(1-\varepsilon)}\left(
\frac{4\pi\mu^2s}{tu}\right)^{\varepsilon}\,
\frac{2(1-\varepsilon+\varepsilon^2)}{(1-\varepsilon)(2-\varepsilon)}
\,\left(1-\varepsilon-\frac{\varepsilon^2 s^2}{2tu}\right) 
\;.
\end{equation}
Needless to say that we have applied also here our ``projector'' 
${\cal F}(p_{\gamma},s_a,s_b)$ of Eq.~(\ref{eq5}) and performed 
the integration over the scalar products involving 
spin vectors according to Eq.~(\ref{eq14}). 

In the final-state collinear case, an expression very similar 
to Eq.~(\ref{subfac1}) is to be used, involving now the
unpolarized quark-to-photon splitting function
\begin{equation} \label{pgamq}
P_{\gamma q} (z) = \frac{1+(1-z)^2}{z}
\end{equation}
and the $2\to 2$ ``pure-QCD'' transversity cross sections in $d$
dimensions, given by:
\begin{eqnarray}
\left\langle \frac{d\delta \hat{\sigma}^{(0)}_{q\bar{q}\to q'\bar{q}'}(s,t,u,
\varepsilon)}{dt} \right\rangle&=&\frac{C_F}{2N_C}\frac{\alpha_s^2}{s^2}
\frac{\mu^{2\varepsilon}}{\Gamma(1-\varepsilon)}\left(
\frac{4\pi\mu^2s}{tu}\right)^{\varepsilon}\,
(2+\varepsilon)\,\frac{t u}{s^2}\;\nonumber \, , \\ 
\left\langle \frac{d\delta \hat{\sigma}^{(0)}_{q\bar{q}\to q\bar{q}}(s,t,u,
\varepsilon)}{dt} \right\rangle  &=&\frac{C_F}{2N_C}\frac{\alpha_s^2}{s^2}
\frac{\mu^{2\varepsilon}}{\Gamma(1-\varepsilon)}\left(
\frac{4\pi\mu^2s}{tu}\right)^{\varepsilon}\,\left[ \,
(2+\varepsilon)\,\frac{t u}{s^2}-\frac{(2-\varepsilon)}{N_C}\frac{u}{s}
\right]\;\nonumber \, , \\
\left\langle \frac{d\delta \hat{\sigma}^{(0)}_{qq\to qq}(s,t,u,
\varepsilon)}{dt} \right\rangle &=&\frac{C_F}{2N_C^2}\frac{\alpha_s^2}{s^2}
\frac{\mu^{2\varepsilon}}{\Gamma(1-\varepsilon)}\left(
\frac{4\pi\mu^2s}{tu}\right)^{\varepsilon}\,
(2-\varepsilon)\; .
\end{eqnarray}
In these expressions, we have neglected contributions $\propto
{\cal O}(\varepsilon^2)$, which do not contribute. Then, the results 
for a fully anticommuting $\gamma_5$ and for the HVBM prescription are 
again the same.

Before coming to our final results, we would like to make two
more comments on the use of our ``projector'' on the azimuthal-angular 
dependence, Eq.~(\ref{eq5}).  In an NLO calculation, carried out in 
$d$ dimensions, we could have a projector that by itself contains
terms $\propto\varepsilon$. Indeed, some of the Born cross sections, when
evaluated in $d$ dimensions, suggest a projector of the form
\begin{equation} 
\label{eq5p}
{\cal F}_{\varepsilon}(p_{\gamma},s_a,s_b)\;=\;
\frac{s}{\pi t u} 
\,\left[ 2 \,(p_{\gamma}\cdot s_a)\, (p_{\gamma}\cdot s_b)\; +\; 
(1-a\varepsilon)\frac{t u}{s} \,(s_a \cdot s_b) \right] \;,
\end{equation} 
with some constant $a$. Clearly, the final answer of the calculation
must not depend on $a$ because our projection is a physical
operation which could be done in experiment.  We have used
the above projector with an arbitrary $a$ and checked that indeed 
no answer depends on $a$.
Also, we have integrated all squared matrix elements over the
spin vectors {\em without} using any projector at all. This
amounts to integrating $\cos(2\Phi)$ over all $0\leq\Phi\leq 2\pi$, 
and, as expected, we get zero in the final answer. It should be
stressed, however, that individual pieces in the calculation 
(the virtual, the $2\to 3$, and the factorization part) do not
by themselves integrate to zero, but only their sum does. In this
way, we have a very powerful check on the correctness of our
calculation.

\subsection{Final Results for Inclusive and Isolated Photon Cross Sections}

For both subprocesses, the final results for the NLO corrections 
can be cast into the following form:
\begin{eqnarray}
&&\hspace*{-1cm}\left\langle 
s\, \frac{d \delta \hat{\sigma}_{ab\to\gamma X}^{(1)}
(s,v,w,\mu_R,\mu_F)}{dvdw} \right\rangle = 
\frac{\alpha\alpha_s(\mu_R)}{\pi^2} 
  \left[ \left(  A_0  \delta (1-w) +   B_0
\frac{1}{(1-w)_+} +  C_0 \right) \ln \frac{\mu_F^2}{s} \right. 
\nonumber \\[3mm]
&&+C_1 \;{\cal I}^{\mathrm{final}}(1-v+v w)\; +
A_2 \delta (1-w) \ln \frac{\mu_R^2}{s} + 
A \delta (1-w)  +B \frac{1}{(1-w)_+} +  C 
\nonumber \\[3mm]
&& + D\left(
\frac{\ln (1-w)}{1-w} \right)_+ + E \ln w + F \ln v 
+ G \ln (1-v) + H \ln(1-w)  + I  \ln (1-vw)\nonumber  \\[3mm]
&&\left. +
J  \ln (1-v+vw) +K \frac{\ln w}{1-w} + L \frac{\ln \frac{1-v}{1-vw}}{1-w}
 +M \frac{\ln (1-v+vw)}{1-w}  \right] \:\:\: ,
\label{final}
\end{eqnarray}
where all coefficients are functions of $v$ and $w$, except those
multiplying the distributions $\delta(1-w)$, $1/(1-w)_+$, 
$\left[ \ln(1-w)/(1-w)\right]_+$ which may be written as 
functions just of $v$. Terms with distributions are present only
for the subprocess $q\bar{q}\to\gamma X$. The coefficients
in Eq.~(\ref{final}) are too lengthy to be given here but are
available upon request.

Let us now specify the function ${\cal I}^{\mathrm{final}}
(z=1-v+vw)$. It results from the 
configurations where the photon is collinear with a final-state
quark or antiquark. As we discussed earlier, these will lead to 
final-state collinear singularities that are absorbed, at the 
factorization scale\footnote{We could also choose a final-state 
factorization scale $\mu_F'\neq \mu_F$ here.} $\mu_F$, into photon 
fragmentation functions. The actual form of ${\cal I}^{\mathrm{final}}$ 
depends on the kind of photon signal under consideration. Let us 
first consider the fully inclusive cross section. In this case,
one just counts all photon candidates in the kinematical bin,
without imposing any constraint on additional particles
in the event. This is the simplest cross section and the one
usually measured in fixed-target experiments. In the theoretical
calculation, final-state singularities arise and there is a 
need to introduce a fragmentation contribution, as discussed 
earlier. 

At collider energies, the background from pions decaying 
into photon pairs is so severe that so-called isolation 
cuts are imposed on the photon. The basic idea is that
photons that have little hadronic energy around them are less 
likely to result from $\pi^0$ decay. The standard procedure
is to define a ``cone'' around the photon by $\sqrt{(\Delta \eta)^2+
(\Delta \phi)^2}\leq R$, where typically $R\approx 0.4 \ldots 0.7$, 
and to demand that the hadronic transverse
energy in the cone be smaller than $\tau\, p_T$, where $\tau$ is a parameter 
of order 0.1. For the theoretical calculation, isolation 
implies a strong reduction of the size of the fragmentation 
contribution because photons produced by fragmentation are
always accompanied by a certain amount of hadronic energy. 
A slightly refined type of isolation has been proposed in 
\cite{ref:frixione}. Again a cone is defined, centered on 
the photon, within which the hadronic transverse energy must not 
exceed the limit $\tau \,p_T$. However, one chooses a larger $\tau \sim 1$ 
and then further restricts the hadronic energy by demanding that for any 
$r\leq R$ the hadronic energy inside a cone of opening $r$
be smaller than roughly $\tau (r/R)^2 p_T$. 
In other words, the closer hadronic energy is deposited
to the photon, the smaller it has to be in order for the
event to pass the isolation cut. This isolation method
has not yet been used in any experiment, but it is possible that 
it will become the choice for the {\sc Phenix} 
experiment at RHIC \cite{ref:phen}. 
On the theoretical side, it has the advantage that it
``eliminates'' any kind of fragmentation contribution \cite{ref:frixione}
because fragmentation is assumed to be a (mainly) collinear process, and 
no hadronic activity is allowed exactly parallel to the photon. 

We recall from the previous section that we have performed
an analytical integration over the full phase space of the
unobserved particles in the final-state. This seems at
first sight to preclude the implementation of an isolation
cut ``afterwards''. However, as was shown in \cite{ref:isol,ref:isol1}, 
it is possible to impose the isolation cut in an
approximate, but accurate, analytical way by introducing
certain ``subtraction cross sections''.  The approximation
is based on assuming the isolation cone to be rather narrow.
In this case, dependence on the cone opening can be shown
to be of the form $a\ln(R)+b+{\cal O}(R^2)$. $a$ and $b$
are straightforwardly determined and yield a very accurate
description of isolation even at $R=0.7$. Analytical 
calculations \cite{ref:isol,ref:isol1} are therefore as capable to 
describe the isolated prompt-photon cross section 
as NLO computations in which phase space integrals 
are performed numerically employing Monte-Carlo techniques 
\cite{ref:mc,ref:frixione,ref:isol1}.

For the cases of the fully-inclusive (``incl.'') cross section,
the standard isolation (``std.''), and for the isolation
proposed in \cite{ref:frixione} (``smooth'') the function 
${\cal I}^{\mathrm{final}}(z=1-v+v w)$ takes the following forms:
\begin{equation}
{\cal I}^{\mathrm{final}}(z)=\left\{
\begin{array}{ll}
P_{\gamma q}(z)\;\ln\left( \frac{\mu_F^2}{s}\right) &
\mbox{incl.} \\ [2mm]
P_{\gamma q}(z)\;\ln\left( \frac{\mu_F^2}{s}\right) 
+\Theta(1-z[1+\tau])\left[ P_{\gamma q}(z)\;\ln\left( 
\frac{(1-z)^2 p_T^2 R^2}{\mu_F^2}\right)+z \right]
&
\mbox{std.} \\ [2mm]
P_{\gamma q}(z)\;\ln\left( \frac{(1-z)^3 p_T^2 R^2}{s\,\tau\,z}
\right) &
\mbox{smooth} \; . 
\end{array}
\right.
\end{equation}
One can see the presence of the quark-to-photon splitting
function $P_{\gamma q}$ of Eq.~(\ref{pgamq}), as is expected
for contributions resulting from near-collinear photon emission
in the final-state. It also becomes clear that for the 
standard isolation the dependence on the final-state factorization
scale is reduced and disappears altogether for the isolation
of \cite{ref:frixione}. This is in line with our remarks above about
the size of the fragmentation contribution in these cases.

\section{Numerical Results}
%
In this Section, we present a first numerical application of our
analytical results. We focus on the main features of
the NLO corrections and describe their impact on the
cross section $d\delta\sigma/dp_T$ and the spin asymmetry 
$A_{\mathrm{TT}}^{\gamma}$. Our predictions will apply
for prompt photon measurements with the {\sc Phenix} detector 
at RHIC. This implies that the pseudorapidity region $|\eta|\le 0.35$
is covered, and only half of the photon's azimuthal angle. Using
Eq.~(\ref{eq2}) we restore the $\cos (2\Phi)$ dependence of
the cross section. We take the two quadrants in $\Phi$ covered by
the {\sc Phenix} detector to be 
$-\pi/4<\Phi<\pi/4$ and $3\pi/4<\Phi<5\pi/4$ and integrate
over these. This gives $\left(\int_{-\pi/4}^{\pi/4}+
\int_{3\pi/4}^{5\pi/4} \right) \cos(2\Phi) d\Phi=2$. 
We consider photons isolated according to the
isolation of \cite{ref:frixione} discussed above,
using $R=0.4$ and $\tau=1$. 

Before we can perform numerical studies of $A_{\mathrm{TT}}^{\gamma}$ we have 
to model the $\delta f$ we will use. Nothing is known experimentally 
about transversity so far. The only guidance is provided by the Soffer 
inequality \cite{ref:soffer}
\begin{equation}
\label{eq:soffer}
2\left|\delta q(x)\right| \leq q(x) + \Delta q(x)
\end{equation}
which gives an upper bound for each $\delta f$.
As in \cite{ref:attlo} we utilize this inequality by saturating 
the bound at some low input scale $\mu_0\simeq 0.6\,\mathrm{GeV}$ using 
the NLO (LO) GRV \cite{ref:grv98} and GRSV (``standard scenario'') 
\cite{ref:grsv} densities $q(x,\mu_0)$ and $\Delta q(x,\mu_0)$,
respectively. For $\mu>\mu_0$ the transversity densities $\delta f(x,\mu)$ 
are then obtained by solving the evolution equations with
the LO \cite{ref:artru,ref:lokernel} or 
NLO \cite{ref:wvkern,ref:nlokernels} kernels.
Obviously, the sign to be used when saturating the inequality is at our
disposal; we choose all signs to be positive. We refer
the reader to \cite{ref:attlo} for more details on our 
model distributions. We note that we will always perform the 
NLO (LO) calculations using NLO (LO) parton distribution functions 
and the two-loop (one-loop) expression for $\alpha_s$. 

\begin{figure}[th]
\vspace*{-0.5cm}
\begin{center}
\epsfig{figure=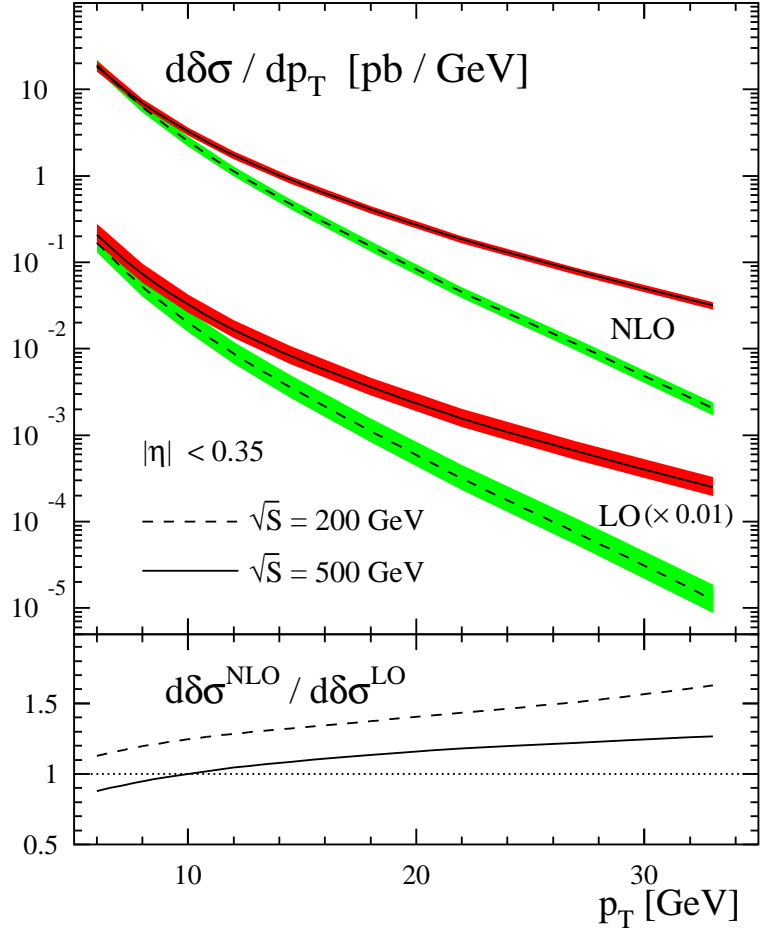,width=0.7\textwidth}
\end{center}
\vspace*{-1.0cm}
\caption{\sf Predictions for the transversely polarized prompt 
photon production cross sections at LO and NLO, for $\sqrt{S}=200$ 
and 500 GeV. The LO results have been scaled by a factor of 0.01.
The shaded bands represent the theoretical uncertainty if
$\mu_F$ $(=\mu_R)$ is varied in the range $p_T/2\le \mu_F \le 2p_T$.
The lower panel shows the ratios of the NLO and LO results for both
c.m.s.\ energies. 
\label{fig:sigma}}
\end{figure}

Figure~\ref{fig:sigma} shows our results for the transversely
polarized prompt photon production cross sections at NLO and LO
for two different c.m.s.\ energies.
The lower part of the figure displays the so called ``$K$-factor''
\begin{equation}
K=\frac{d\delta\sigma^{\rm NLO}}{d\delta\sigma^{\rm LO}} \;\; .
\end{equation}
One can see that NLO corrections are somewhat smaller for
$\sqrt{S}=500\,\mathrm{GeV}$ and increase with $p_T$.
As we have mentioned in the Introduction, one reason why it is 
generally important to know NLO corrections is that they 
should considerably reduce the dependence of the cross
sections on the unphysical factorization and renormalization scales. 
In this sense, the $K$-factor has actually limited significance 
since it is likely to be rather scale dependent through the presence
of the LO cross section in its denominator. The improvement in
scale dependence when going from LO to NLO is, therefore, a 
better measure of the impact of the NLO corrections.
The shaded bands in the upper panel of Fig.~\ref{fig:sigma}
indicate the uncertainties from varying the scales 
in the range $p_T/2 \leq \mu_R=\mu_F \leq 2 p_T$. The solid and dashed
lines are always for the choice where all scales are set to $p_T$, 
and so is the $K$ factor underneath. One can see that the scale 
dependence indeed becomes much weaker at NLO.
%

%
\begin{figure}[th]
\vspace*{-0.5cm}
\begin{center}
\epsfig{figure=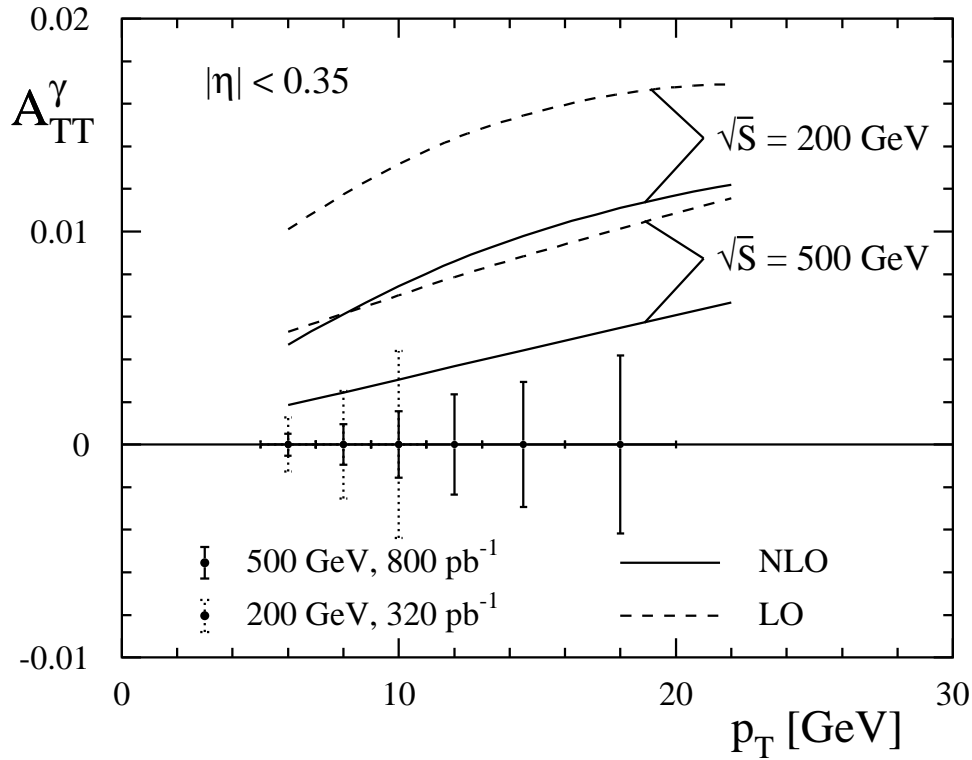,width=0.85\textwidth}
\end{center}
\vspace*{-1cm}
\caption{\sf Predictions for the transverse spin asymmetry 
$A_{\mathrm{TT}}^{\gamma}$ 
for isolated prompt photon production in LO and NLO for
$\sqrt{S}=200$ and 500 GeV. The ``error bars'' indicate the
expected statistical accuracy for bins in $p_T$ (see text).
\label{fig:att}}
\end{figure}
Figure~\ref{fig:att} shows the spin asymmetry $A_{\mathrm{TT}}^\gamma$ 
which is perhaps the main quantity of interest here,
calculated at LO\footnote{We note that our LO asymmetries are
larger than those reported in \cite{ref:attlo}. This is due
to an error in the numerical computation in \cite{ref:attlo}. Our
LO curves in Fig.~2 correct this mistake.} 
(dashed lines) and NLO (solid lines).
We have again chosen all scales to be $p_T$. Due to a larger
$K$ factor for the unpolarized cross section, the 
asymmetry is smaller at NLO than at LO. We also display
in Fig.~\ref{fig:att} the statistical errors expected in experiment.
They may be estimated by the formula \cite{ref:rhic}
\begin{equation} \label{error}
\delta A_{\mathrm{TT}}^{\gamma} 
\simeq \frac{1}{P^2\sqrt{{\cal L}\sigma_{\rm bin}}} \; ,
\end{equation}
where $P$ is the transverse 
polarization of each beam, ${\cal L}$ the integrated
luminosity of the collisions, and $\sigma_{\rm bin}$ the unpolarized
cross section integrated over the $p_T$-bin for which the error is to be
determined. We have used $P=0.7$ and ${\cal L}=320 (800)$/pb for
$\sqrt{S}=200(500)$ GeV. 

\section{Conclusions}
%

We have presented in this paper the complete NLO QCD corrections for the 
partonic hard-scattering cross sections relevant for the
spin asymmetry $A_{\mathrm{TT}}^{\gamma}$ for high-$p_T$ prompt photon 
production in transversely polarized proton-proton collisions. This asymmetry 
could be a tool to determine the transversity content of the nucleon at RHIC.

Our calculation is based on a largely analytical evaluation 
of the NLO partonic cross sections. 
We have presented a simple technique for treating, in an NLO calculation,
the azimuthal-angle dependence introduced by the transverse spin vectors.  
We will apply this technique to other $A_{\mathrm{TT}}$ in the 
future, such as for inclusive pion and jet production \cite{ref:msv}. 

We found that at RHIC energies the NLO corrections to the polarized cross 
section are somewhat smaller than those in the unpolarized case. The
transversely polarized cross section shows a significant reduction of scale
dependence when going from LO to NLO. 

\section*{Acknowledgments}
%
We are grateful to R.L.\ Jaffe for interesting discussions.
M.S.\ thanks the RIKEN-BNL Research Center and Brookhaven National Laboratory 
for hospitality and support during the final steps of this work and 
A.\ Sch\"{a}fer for discussions.  
W.V.\ is grateful to RIKEN, Brookhaven National Laboratory and the U.S.\
Department of Energy (contract number DE-AC02-98CH10886) for
providing the facilities essential for the completion of this work.
This work is supported in part by the ``Bundesministerium f\"{u}r Bildung und
Forschung (BMBF)'' and the ``Deutsche Forschungsgemeinschaft (DFG)''.

\section*{Appendix: NLO Transversity Drell-Yan Cross Section
with Projection Technique}

In this Appendix we briefly report the results we find 
for the NLO corrections to the Drell-Yan ``coefficient function''
$\delta C^{\mathrm{DY}}$ when using our projection method of Sec.~2.2. 
For details on the kinematics for the process, see 
\cite{ref:dynlo,ref:wvkern}. 
We use a fully anticommuting $\gamma_5$ and choose the scales
$\mu_F=\mu_R=Q$ everywhere, with $Q$ the dilepton mass. The
LO cross section and the virtual corrections at NLO rely
on the underlying $2\to 2$ reaction $q\bar{q}\to l^+ l^-$. The
real-emission NLO $2\to 3$ process is $q\bar{q}\to l^+ l^- g$.
We apply our projector, Eq.~(\ref{eq5}), to the
squared matrix elements for each of these processes and integrate
over the appropriate phase spaces. For the $2\to 3$ process this
gives:
\begin{eqnarray}
\delta C^{\mathrm{DY}}_{2\to 3}&=&\frac{\alpha_s}{2\pi}
\frac{C_F(4\pi)^{2\varepsilon}}{\Gamma(1-2\varepsilon)}
\left[ \left( \frac{2}{\varepsilon^2} + \frac{13}{3 \varepsilon}
- \frac{\pi^2}{3} -\frac{29}{18}\right)\delta (1-z) +
\left( -\frac{4}{\varepsilon}-\frac{26}{3}\right) \frac{z}{(1-z)_+}\right.
\nonumber \\
&&\left.+ 8 z\left( \frac{\ln (1-z)}{1-z} \right)_+
- 4 z \frac{\ln z}{1-z}- 6 z \frac{\ln^2 z}{1-z} + 4 (1-z) \right]  \; ,
\end{eqnarray}
where $z=Q^2/s$. For the virtual 
contributions we get
\begin{equation}
\delta C^{\mathrm{DY}}_{\mathrm{virt.}}=\frac{\alpha_s}{2\pi}
\frac{C_F(4\pi)^{2\varepsilon}}{\Gamma(1-2\varepsilon)}\left[ 
-\frac{2}{\varepsilon^2} - \frac{22}{3 \varepsilon}
+ \pi^2 -\frac{116}{9}
\right]\delta (1-z)  \; ,
\end{equation}
and for the $\overline{\rm{MS}}$ collinear-factorization term
\begin{equation}
\delta C^{\mathrm{DY}}_{\mathrm{fact.}}=\frac{\alpha_s}{2\pi}
\frac{C_F(4\pi)^{2\varepsilon}}{\Gamma(1-2\varepsilon)}\left[ 
\left( \frac{3}{\varepsilon}+\frac{13}{2} \right) \delta (1-z)
+ \left( \frac{4}{\varepsilon}+\frac{26}{3}\right) \frac{z}{(1-z)_+}
\right]  \; .
\end{equation}
Adding all terms, the poles cancel, and one obtains the
NLO $\overline{\rm{MS}}$ coefficient function:
\begin{eqnarray}
\delta C^{\mathrm{DY}}(z)&=&\frac{\alpha_s}{2\pi}C_F 
\left[ \left(  \frac{2}{3} \pi^2 -8\right)\delta (1-z) 
+ 8 z\left( \frac{\ln (1-z)}{1-z} \right)_+
- 4 z \frac{\ln z}{1-z}- 6 z \frac{\ln^2 z}{1-z} + 4 (1-z) \right]
\nonumber \\ &&
\end{eqnarray}
in agreement with \cite{ref:wvkern}.

%

%
\end{document}